# Synthesis of titanate nanostructures using amorphous precursor material and their adsorption/photocatalytic properties

E.K. Ylhäinen[1], M.R. Nunes[2], A.J. Silvestre[3] and O.C. Monteiro[4*]

[1] *Department of Chemistry and Bioengineering, Tampere University of Technology, P.O. Box 541, 33101 Tampere, Finland*

[2] *University of Lisbon, Faculty of Sciences, Department of Chemistry and Biochemistry and CCMM, 1749-016 Lisboa, Portugal*

[3] *Instituto Superior de Engenharia de Lisboa and ICEMS, R. Conselheiro Emídio Navarro 1, 1959-007 Lisboa, Portugal*

[4] *University of Lisbon, Faculty of Sciences, Department of Chemistry and Biochemistry and CQB, 1749-016 Lisboa, Portugal*

## Abstract

This paper reports on a new and swift hydrothermal chemical route to prepare titanate nanostructures (TNS) avoiding the use of crystalline $TiO_2$ as starting material. The synthesis approach uses a commercial solution of $TiCl_3$ as titanium source to prepare an amorphous precursor, circumventing the use of hazardous chemical compounds. The influence of the reaction temperature and dwell autoclave time on the structure and morphology of the synthesised materials was studied. Homogeneous titanate nanotubes with a high length/diameter aspect ratio were synthesised at 160 ºC and 24 h. A band gap of 3.06±0.03 eV was determined for the TNS samples prepared in these experimental conditions. This value is red shifted by 0.14 eV compared to the band gap value usually reported for the $TiO_2$ *anatase*. Moreover, such samples show better adsorption capacity and photocatalytic performance on the dye rhodamine 6G (R6G) photodegradation process than $TiO_2$ nanoparticles. A 98% reduction of the R6G concentration was achieved after 45 minutes of irradiation of a 10 ppm dye aqueous solution and 1g $L^{-1}$ of TNS catalyst.

* Corresponding author: Phone:+351 217500865; Fax:+351 217500088; E-mail: ocmonteiro@fc.ul.pt

## 1. Introduction

Tubular nanostructures have become one of the most important research subjects in nanotechnology. Among the nanotubular materials that have been synthesised during the last two decades, titanate nanostructures (TNS) have attracted increasing attention in recent years [1-3]. By combining properties of conventional $TiO_2$ nanoparticles with the properties of layered titanates, TNS have potential applications covering a wide variety of fields including photocatalysis [4], as substrate to decorate with different active catalysts [5], dye-sensitized solar cells [6], transparent optical devices [7], and gas or humidity sensors [8].

Since the discovery of the $TiO_2$ *anatase*-based alkaline hydrothermal chemical route to prepare TNS by Kasuga et al. [9], the mechanism of formation and the method for regulating their morphology (length/diameter aspect ratio, wall thickness, size distribution) and stability are at present the subject of intense research [1,10-13]. It has been demonstrated that all main $TiO_2$ polymorphs can be used as precursors in the synthesis of such nanostructures [1,2]. Titanate nanorods have been also synthesized using sodium titanate [14] or using an amorphous $TiO_2.nH_2O$ gel as raw materials [15]. Despite the several techniques available to grow $TiO_2$ nanostructures, *e.g.* template methods, electrochemical anodic oxidation, spin-on processes, metalorganic chemical vapour deposition and pyrolysis routes [1,2,15], until now the hydrothermal chemical routes are still considered to be the most effective and low cost method to prepare titanate nanotubular structures, most of the reported procedures making use of crystalline $TiO_2$ as precursor material [1,2,16]. However, the use of either distinct commercial sources of $TiO_2$ powders, or of those synthesised in research laboratories frequently leads to nanostructures with different microstructural features, [1,2,16,17] rendering the reproducibility of the TNS synthesis process highly dependent on the $TiO_2$ starting material and far from being well established. Moreover, it is well known that due to its catalytic properties $TiO_2$ has been highly investigated as a promising photocatalyst for the treatment of industrial wastewaters and polluted air, since it has a good photoactivity under ultraviolet light irradiation, is non-toxic, water insoluble, and comparatively inexpensive [18,19]. However, the major drawbacks for its wide practical application in this field it is its high charge recombination rate and its wide band gap (*anatase*, ca. 3.2 eV) which limits the photogeneration of electrons and holes to the UV light below 387 nm (only ~5% of the solar radiation reaching the Earth) [19]. Therefore, the synthesis of a $TiO_2$-based material either with a broader range of light absorption and/or a lower charge recombination rate would be an important step toward the development of a higher efficient photoactive material.

In this work a new hydrothermal chemical route is proposed to prepare homogenous and stable titanate nanostructures. The process is simple, highly reproducible and avoids the use of crystalline $TiO_2$ as precursor material. The synthesis approach uses a commercial solution of $TiCl_3$ as titanium source to prepare an amorphous precursor, circumventing the use of hazardous chemical compounds. The influence of the experimental parameters on the structure and morphology of the titanate nanostructures prepared was studied. Results on the adsorption and photocatalytic response of the TNS synthesised materials are also presented.

## 2. Materials and Methods

All reagents were of analytical grade (Aldrich and Fluka) and were used as received. The solutions were prepared with Millipore Milli-Q ultra pure water.

### *2.1. Materials*

#### *2.1.1. TNS precursor synthesis*

The TNS precursor was prepared using a procedure reported previously [20]. A titanium trichloride solution (10 wt.% in 20-30 wt.% HCl) diluted in a ratio of 1:2 in standard HCl solution (37 %) was used as titanium source. To this solution a 4 M ammonia aqueous solution was added dropwise under vigorous stirring until complete precipitation of a white solid. The resulting suspension was kept at rest for 15 h at room temperature and then filtered and vigorously rinsed with deionised water in order to remove the remaining ammonium and chloride ions.

### *2.1.2 TNS synthesis*

The synthesis of the TNS samples was performed in an autoclave system using 3 g of precursor in ca. 30 ml of NaOH 10 M aqueous solution. Samples were prepared using temperatures ranging from 130 ºC to 220 ºC, and autoclave dwell times (*t*) varying between 12 h and 72 h. After being washed with water until pH 7, the solids were dried and stored. A control sample was prepared at room temperature during 72 hours. For comparative purposes crystalline *anatase* $TiO_2$ nanoparticles were also prepared by the above procedure, using water instead of the NaOH solution as solvent during the hydrothermal treatments. For the sake of clarity, the titanate nanostructured samples were labelled as "TNS'temperature'time" and the $TiO_2$ nanoparticles samples as "NP'temperature'time"; e.g. the TNS16024 and NP16024 samples are the titanate and $TiO_2$ nanoparticles samples prepared at 160 ºC during 24 h, respectively.

### *2.2 Photodegradation experiments*

All the photodegradation experiments were conducted using a 250 ml refrigerated photoreactor [21]. The radiation source used was a 450 W Hanovia medium-pressure mercury-vapour lamp, the total irradiated energy being 40-48% in the ultraviolet range and 40-43% in the visible region of the electromagnetic spectrum. Suspensions were prepared by adding 15 mg of powder to 150 ml of 10 ppm rhodamine 6G (R6G) aqueous solution. Prior to irradiation, suspensions were stirred in darkness for 60 min to ensure adsorption equilibrium. During irradiation, suspensions were sampled at regular intervals, centrifuged and analysed by UV-Vis spectroscopy.

### *2.3 Characterization*

X-ray powder diffraction was performed using a Philips X-ray diffractometer (PW 1730) with automatic data acquisition (APD Philips v3.6B), using Cu Kα radiation ($\lambda$ = 0.15406 nm) and working at 40 kV/30 mA. The diffraction patterns were collected in the $2\theta$ range of 7º - 60º with a 0.02º step size and an acquisition time of 2.0 s/step. The diffractometer was calibrated before every measurement. FTIR analyses were conducted with a FTIR Nicolet F-6700 spectrophotometer, using KBr pellets. Transmission electron microscopy (TEM), high resolution transmission electron microscopy (HRTEM) and selected area electron diffraction (SAED) were carried out on a JEOL 200CX microscope operating at 200 kV. The X-ray photoelectron spectroscopy (XPS) spectra were taken in CAE mode (30 eV), using an Al (non-monochromate) anode. The accelerating voltage was 15 kV. The quantitative XPS analysis was performed using the Avantage software. The optical characterization of the samples was carried out by UV-Vis diffuse reflectance using a Shimadzu UV-2450PC spectrometer. The diffuse reflectance spectra (DRS) were recorded in the wavelength range of 220-850 nm. Specific surface areas were obtained by the BET method, from nitrogen (Air Liquide, 99.999%) adsorption data at −196 °C, using a volumetric apparatus from Quantachrome mod NOVA 2200e. The samples, weighing between 40 and 60 mg, were previously degassed for 2.5 h at 150 °C at a pressure lower than 0.133 Pa. An UV-Vis spectrophotometer Jasco V560 was used for monitoring the absorption of R6G solutions.

## 3. Results and discussion

### *3.1. Precursor characterization*

Before the TNS synthesis, the characterization of the TNS precursor material was carried out in order to study its structure and composition. Figure 1 shows a typical XRD pattern obtained for the precursor material. As can be seen no diffraction peaks were obtained, this result suggesting a high degree of amorphization of this material. The inset of figure 1 shows a TEM image of the precursor morphology, revealing nanometric size particles (~5 nm) with unclear boundaries and strongly aggregated. The amorphous nature of the TNSs precursor material was confirmed by SAED analyses taken over different zones of the studied material (see SAED image overlapping the TEM image of figure 1).

The XPS analysis was performed in order to infer the elemental composition and electronic state of the precursor sample. Figure 2 shows the Ti 2p and O 1s survey spectra of the precursor material. The Ti 2p1/2 and Ti 2p3/2 spin-orbital splitting photoelectrons are located at binding energies of 467.81 eV and 462.20 eV, respectively. The peak separation of 5.61 eV between the Ti 2p1/2 and Ti 2p3/2 signals is in agreement with other reported values for $TiO_2$ structures [22,23]. The wide and asymmetric peak of O 1s spectra indicate that there would be more than one chemical state according to the binding energy. The O 1s resolved signal resulted in three peaks located at a binding energy of 535.2 eV, 533.27 eV and 531.58 eV. The main peak at 533.27 eV can be assigned to Ti–O–Ti bonds. The peaks located at 535.2 eV and 531.58 eV can be attributed to the hydroxyl species and to the existence of some carbon contamination, respectively. The binding energy difference of 71.07 eV between the observed peaks positions of Ti 2p1/2 and O 1s (oxide) is also in agreement with $TiO_2$ reported values [22]. However, the titanium-to-oxygen ratio determined by integrating the areas under the Ti 2p and O 1s peaks is 0.25:1, which suggests a different stoichiometry that the one of $TiO_2$.

### *3.2. TNS materials characterization*

The TNS nanostructures were first prepared at 160 ºC during 24 hours. As can be seen in the corresponding diffractogram of figure 3, the diffraction peaks at $2\theta$ = 10.6º, 24.5º, 28.6º and 48.6º could be clearly identified and are in agreement with a titanate layered structure. The broad peaks at $2\theta = 10.6º$ and 28.6º correspond to interlayer spacing in layered titanates. The peaks at $2\theta = 24.5º$ and 48.6º indicate the presence of hydrogen and sodium trititanates. Considering these facts, the XRD patterns obtained cannot thoroughly confirm the structure of the TNSs, but are in perfect agreement with previous studies and are assumed to be nanocrystalline $Na_xH_{2-x}Ti_3O_7$ [24-27]. The presence of the elements Na, Ti and O were also confirmed by EDS analyses (not shown). Figure 4 shows the diffractogram of $TiO_2$ nanoparticles sample also prepared at 160 ºC during 24 hours. As can be seen, the pattern matches the $TiO_2$ *anatase* phase JCPDS-ICDD file No. 21-1272, confirming that only one crystalline phase was formed during the synthesis process. TEM analyses allowed to estimate the $TiO_2$ nanoparticles grain size ranging between 20 and 30 nm (see inset of figure 4).

### *3.2.1. Reaction time influence*

In order to study the influence of reaction time on the crystal structure and morphology of the synthesized materials, TNS samples were prepared at 160 °C for autoclave dwell times of 12, 24, 36, 48 and 72 h. Their XRD patterns are shown in figure 3. As can be seen by increasing the duration of the hydrothermal treatment, the XRD peaks become narrower and more intense. Notice that for samples prepared using $t < 12$ h no crystalline material was obtained (not shown). The XRD patterns of samples prepared for 36, 48 and 72 h are similar. However these samples have a distinct structural feature compared to those prepared for 12 and 24 h: by increasing the reaction time from 24 to 36 h and above, the diffraction peaks are well defined, at slightly different angles and new peaks are detected. The broad peak at $2\theta \sim 10º$ is shifted to higher values which indicate a decreasing of the interlayer spacing [25]. However the peak position shifts again to lower angles after reaching its maximum at heating time of 36 hours. The increased interlayer spacing may occur due to adsorbed water molecules. In fact, adsorbed water was identified in the TNS samples by FTIR analyses. A strong band at 3400 $cm^{-1}$ corresponding to the O-H stretching vibration and a band at 1630 $cm^{-1}$ corresponding to the H-O-H bending vibration were clearly identified (not shown).

The influence of the hydrothermal reaction time on the samples structure can be seen in more detail by comparing the XRD patterns of the TNS16024 sample with the materials prepared using times up to 48 h. As revealed by the TNS16024 and TNS16048 diffractograms in figure 3, increasing autoclave dwell time up to 48 h the crystalline structure of the TNS16048 sample can be indexed to both the $H_2Ti_3O_7$ and the $Na_2Ti_3O_7$ JCPDS-ICDD files nº 41-192 and 31-1329, respectively. Note that these XRD results cannot unambiguously confirm the structure of the synthesised TNS samples, but are consistent with the formation of $Na_xH_{2-x}Ti_3O_7$ complex titanates, in agreement with other reported studies [25,26].

The morphology of the different samples prepared at 160 ºC was analyzed by TEM/HRTEM. From the micrographs shown in figure 5 it can be concluded that the reaction time has a strong influence on the powders morphology. As can be seen, the sample synthesized with $t = 24$ h (TNS16024) is very homogeneous and consists of thin tubular nanostructures with a high length/diameter aspect ratio (figure 5a) while the sample prepared at $t = 36$ h (TNS16036) shows a mixture of nano and micro scale structures (figure 5b). The sample prepared at $t = 48$ h (TNS16048) is characterized by broomstick-like large bundles (figure 5c). Identical morphologies have been reported for $H_2Ti_3O_7$ nanorods prepared via alkaline hydrothermal route using an amorphous $TiO_2.nH_2O$ gel as a precursor material at 200ºC during 20 hours [15]. Increasing further the reaction time up to 72 h (TNS16072), the broomstick-like large bundles lead to titanate nanostructures, yet with a larger diameter than for the samples prepared at $t = 24$ h (figure 5d). An interlayer distance of ca. 0.8 nm was estimated for the TNS16072 sample by a HRTEM image (inset of figure 5d). This result is in agreement with previously reported results [27].

### *3.2.2. Reaction temperature influence*

The influence of the hydrothermal reaction temperature on the synthesized nanomaterial properties was evaluated by studying samples prepared at 130, 160, 200 and 220 ºC for autoclave dwell time of 24 h. All samples were identified as sodium and hydrogen tri-titanate nanotubes, the predominant crystalline phase being $Na_2Ti_3O_7$. The XRD patterns of these samples are shown in figure 6. As can be seen, by increasing the temperature the diffraction peak related with the interlayer spacing ($2\theta \sim 10^o$) is narrowing and shifting to higher angles, changing for instance from 10.06º to 10.7º as the reaction temperature varies from 130 ºC to 220 ºC. This interlayer spacing decrease can be correlated with the release of structural water, taking place at high temperatures [28]. Decreasing interlayer distance could also indicate transformation from tube to a cylindrical wire or a nanorod structure.

In what concerns the morphology of the samples, for reaction temperatures varying between 130 ºC and 160 ºC no significant differences were observed in the length/diameter aspect ratio of the samples, all of them with very similar morphologies to that shown in figure 5a. However, for higher reaction temperatures an increase of both length and diameter of the TNS structures with increasing temperature was observed. Figure 7 shows HRTEM images of the samples prepared at 200 ºC and 220 ºC for 24 h. Larger diameters, in the range 75-100 nm, were obtained for samples synthesized at 220 ºC, comparatively to the samples prepared at 200ºC.

The overall effects of reaction time and temperature can be analyzed in more detail by comparing the XRD patterns of TNS16072 and TNS20024 samples shown in figure 8. The patterns are similar and most of the diffraction peaks can be indexed to the crystalline structures of $H_2Ti_3O_7$ and $Na_2Ti_3O_7$. Nevertheless some diffraction reflections cannot be indexed within those two crystalline phases. In previous studies it has been suggested that $H_2Ti_3O_7$ nanotubes can start to transform into $TiO_2$ *anatase* at temperatures over 160 ºC [29,30]. Concurrently and even considering that sodium stabilizes the nanotubular structure, the $Na_2Ti_3O_7$ can also be transformed into sodium hexa-titanate ($Na_2Ti_6O_{13}$). Our results are in agreement with these facts: the TNS16072 and TNS20024 XRD patterns are consistent with the existence of $H_2Ti_3O_7$ and $Na_2Ti_3O_7$ as predominant crystalline phases and small amounts of $TiO_2$ *anatase* and $Na_2Ti_6O_{13}$.

### *3.2.3. UV-Vis photo-response*

Due to the relevance of $TiO_2$ and related structures on photo-irradiated applications, the UV-Vis absorption behaviour of TNS16024 sample was analysed and compared with the photo-response of a $TiO_2$ nanoparticles sample prepared using the same time and temperature conditions (sample NP16024). Optical characterization of both samples was carried out by measuring the diffuse reflectance, $R$, at room temperature. $R$ is related with the Kubelka-Munk function $F_{KM}$ by the relation $F_{KM}(R) = (1-R)^2/2R$ [31]. The optical band gap energy of the TNS sample was calculated by plotting the function $f_{KM} = (F_{KM}\, h\upsilon)^{0.5}$ *vs.* energy (Tauc plot), where $h$ stands for the Planck constant and $\upsilon$ for the frequency. The linear part of the curve was extrapolated to $f_{KM} = 0$ to get the indirect band gap energy of the sample. Figure 9a shows the diffuse reflectance spectra of both samples. As can be seen, the TNS sample absorbs more radiation in the visible range than the $TiO_2$ *anatase* nanoparticles sample, the TNS sample absorption band edge being slightly shifted to the higher wavelength region relative to that of NP sample. Figure 9b shows the Tauc plot of sample TNS10624, from which an indirect band gap of 3.06±0.03 eV was inferred. This value is red shifted by 0.14 eV compared to the band gap value usually reported for the $TiO_2$ *anatase* and it is very close to the reported band gap values of 3.1 eV for titanate nanotubes [32] and 3.08 eV for titanate nanoribbons [33].

### *3.2.4. Dye Adsorption and R6G photocatalytic degradation*

Based on the visible photo-response results, the photocatalytic activity of the TNS16024 sample was evaluated in the cationic R6G photodegradation process. The NP16024 sample was also used for comparative proposes. Prior to the irradiation experiments, the surface area and the dye adsorption ability were analysed. The BET surface area calculated for the TNS16024 sample was 240.24 $m^2/g$, while for NP16024 sample a much lower value of 67.50 $m^2/g$ was obtained.

The R6G adsorption and photodegradation profiles are shown in figure 10. Note that the negative time was used for convenience to show the dye adsorption extension before irradiation. Concerning the adsorption process, 79.9% of the R6G was adsorbed in the TNS16024 surface after 1 h of dark stirring conditions, this result being a clear evidence of the high R6G adsorption capability of this material. Identical adsorbability values were reported by Huang et al. [34] for methylene blue onto titanate nanotubes with similar BET surface areas, prepared via hydrothermal treatment of $TiO_2$ *anatase* in NaOH 4 M aqueous solution during 96 hours at 180 ºC. Note, however, that only 16.1% of the dye was adsorbed by the $TiO_2$ nanoparticles sample. If one considers that the adsorption is proportional to the surface area then, taking into account the TNS16024 and NP16024 BET surface areas and the adsorbed amount of dye by the $TiO_2$ nanoparticles, only 57.3% of R6G would be expected to be adsorbed by the TNS sample. Therefore in order to explain the higher experimental TNS adsorption results not only the surface area factor should be considered but also the TNS high ion-exchange ability and the dye cationic character.

For the irradiation period, the profiles of figure 10 allow to conclude that the TNS16024 sample shows photocatalytic activity in the R6G photodegradation process. As can be seen, without catalyst (photolysis) a decrease of about 60% of the R6G concentration was achieved after 45 min of irradiation. Using TNS16024 as catalyst, a R6G concentration reduction of

~98% was achieved, for the same period of time. Only a decrease of ~84% was attained using the NP16024 nanoparticles sample. Thus, concerning the R6G photodegradation reaction, the TNS material presents better catalytic performance than the $TiO_2$ nanoparticles material. These results could be supported not only by the red shift inferred for the TNS sample optical band gap, and thus the increase of the energy absorption in the visible range, but also by a possible lower recombination rate for the titanate nanotubes compared to that of $TiO_2$ nanopowders. It should be noted that, even taking into consideration these promising results, after complete solution decolourization more ~60 min were required for the complete catalyst recovering.

## 4. Conclusions

A swift hydrothermal chemical route to prepare homogenous and stable titanate nanostructures avoiding the use of crystalline $TiO_2$ as precursor material was proposed. The influence of the reaction temperature and dwell autoclave time on the structure and morphology of the synthesised materials was studied. The best experimental conditions to prepare the titanate nanotubular structures were set at 160 ºC and 24 h. In these experimental conditions very homogeneous nanotubular structures with a high length/diameter aspect ratio can be synthesised. An indirect optical band gap energy of 3.06±0.03 eV was determined for such nanostructures, this value being red shifted by 0.14 eV when compared to that of $TiO_2$ *anatase*.

In what concerns the R6G photodegradation process, the TNS material prepared in the above referred experimental conditions exhibits better photocatalytic performance than the $TiO_2$ nanoparticles prepared using identical conditions. The enhanced photocatalytic activity of the titanate nanostructures may be associated to the increase on the light absorption in the visible region range of the TNSs when compared with the $TiO_2$ nanocrystalline material.

## Acknowledgments

This work was supported by Fundação para a Ciência e Tecnologia (PTDC/CTM-NAN/113021/2009). E. K. Ylhäinen acknowledges IAESTE for a grant. The authors thank P.I. Teixeira for a critical reading of the manuscript.

**Figure captions**

**Figure 1** - XRD pattern of the TNS precursor material. The inset shows a TEM and SAED images of a TNS precursor sample. The SAED image was taken over the region delimited by the dashed white square shown in the TEM micrograph.

**Figure 2** - a) Ti 2p and b) O 1s XPS survey spectra of the TNS precursor material.

**Figure 3** - XRD patterns of the titanate nanostructures prepared at 160 ºC and different reaction times: 12 h (TNS16012), 24 h (TNS16024), 36 h (TNS16036), 48 h (TNS16048) and 72 h (TNS16072). Symbols: (∗) - $Na_2Ti_3O_7$, JCPDS-ICDD file No. 31-1329, (•) - $H_2Ti_3O_7$, JCPDS-ICDD file No. 41-192.

**Figure 4** - XRD pattern of the $TiO_2$ nanoparticles sample NP16024. All the diffraction lines match the $TiO_2$ anatase phase JCPDS-ICDD file No. 21-1272. The inset shows a TEM micrograph of the same sample.

**Figure 5** - TEM images of titanate nanostructures prepared at 160 ºC for a) 24 h, b) 36 h, c) 48 h and d) 72 h.

**Figure 6** - XRD patterns of the titanate nanostructures synthesized for 24 h at different temperatures. Symbols: (∗) - $Na_2Ti_3O_7$, JCPDS-ICDD file No. 31-1329, (•) - $H_2Ti_3O_7$, JCPDS-ICDD file No. 41-192.

**Figure 7** - TEM micrographs of TNS samples prepared at a) 200 ºC and b) 220 ºC for an autoclave dwell time of 24 h.

**Figure 8** - XRD patterns of the samples prepared at 200 ºC and 24 h (TNS20024) and at 160 ºC and 72 h (TNS16072). The vertical lines are assigned to the $TiO_2$ *anatase* phase diffraction

pattern according to the JCPDS-ICDD file No. 21-1272.

**Figure 9** - a) Diffuse reflectance spectra of the titanate TNS16024 (solid line) and nanoparticles NP16024 (dashed line) samples, prepared at 160 ºC for 24 hours. b) Tauc plot of the TNS16024 sample from which an indirect band gap of 3.06±0.03 eV was determined.

**Figure 10** - Photocatalytic decolourization of a 10 ppm R6G aqueous solution (150 ml) using 15 mg of catalyst. $C_i$ and $C_t$ stands for the initial and over time concentrations of R6G, respectively.

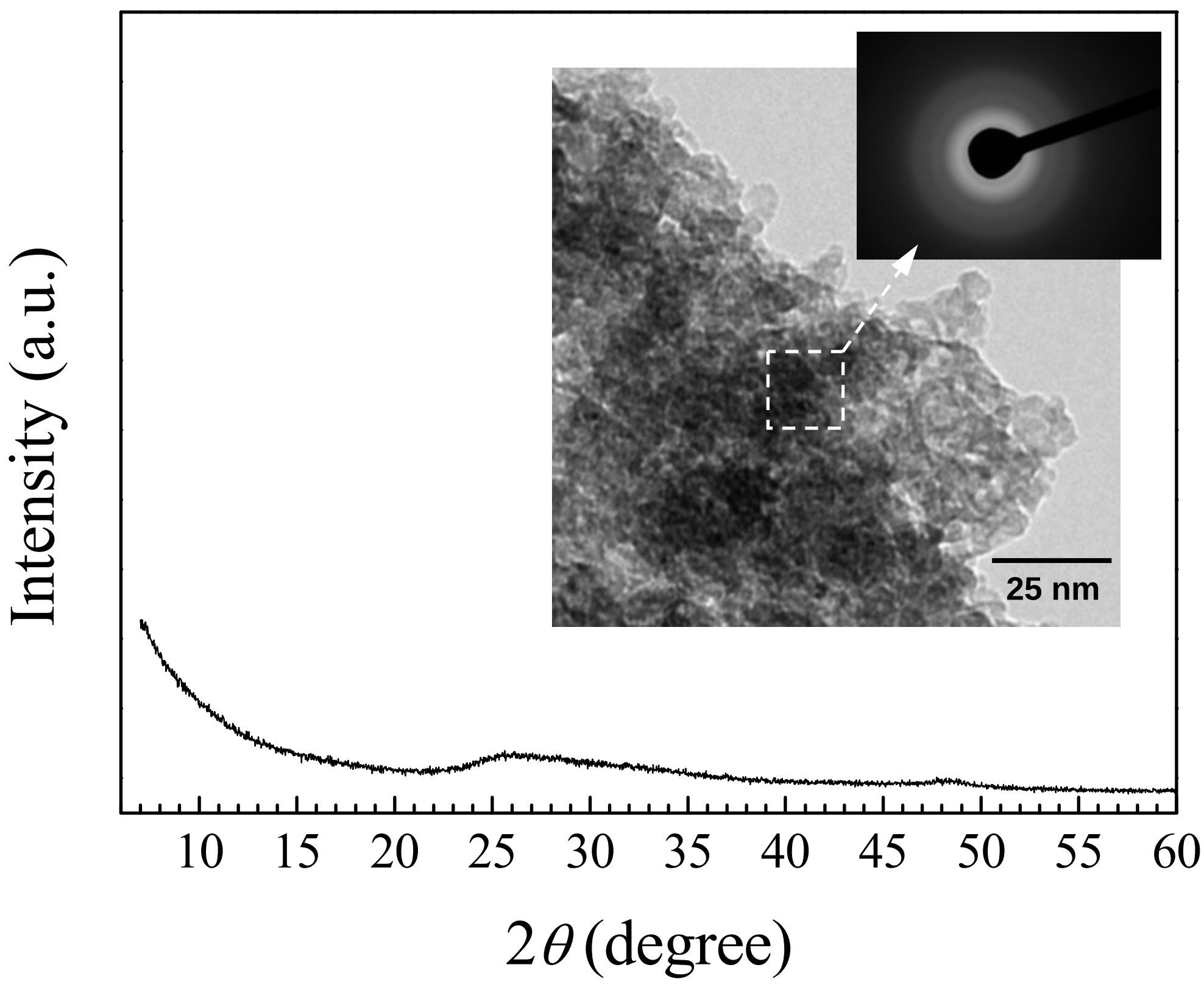

**Figure 1** - XRD pattern of the TNS precursor material. The inset shows a TEM and SAED images of a TNS precursor sample. The SAED image was taken over the region delimited by the dashed white square shown in the TEM micrograph.

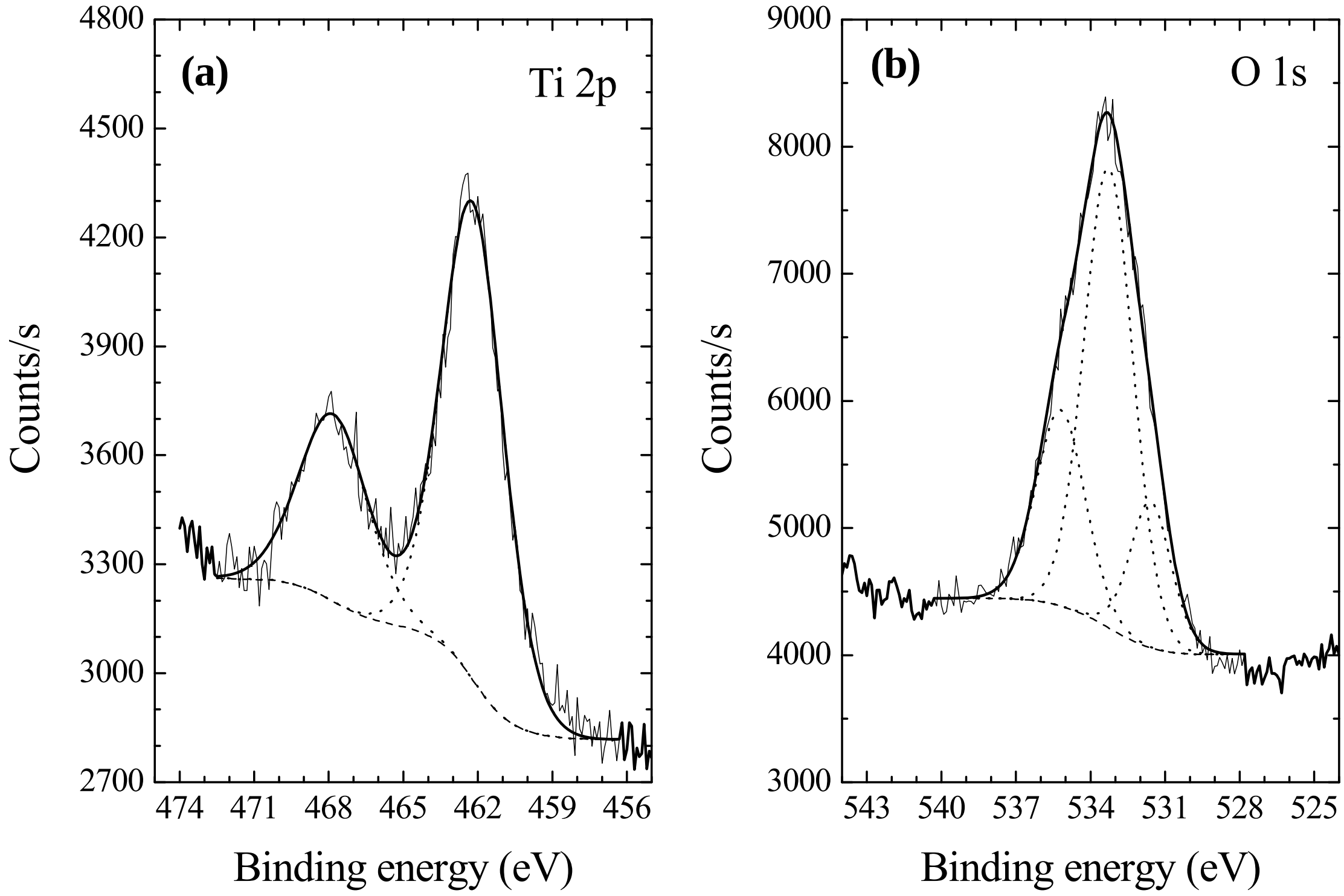

**Figure 2** - a) Ti 2p and b) O 1s XPS survey spectra of the TNS precursor material.

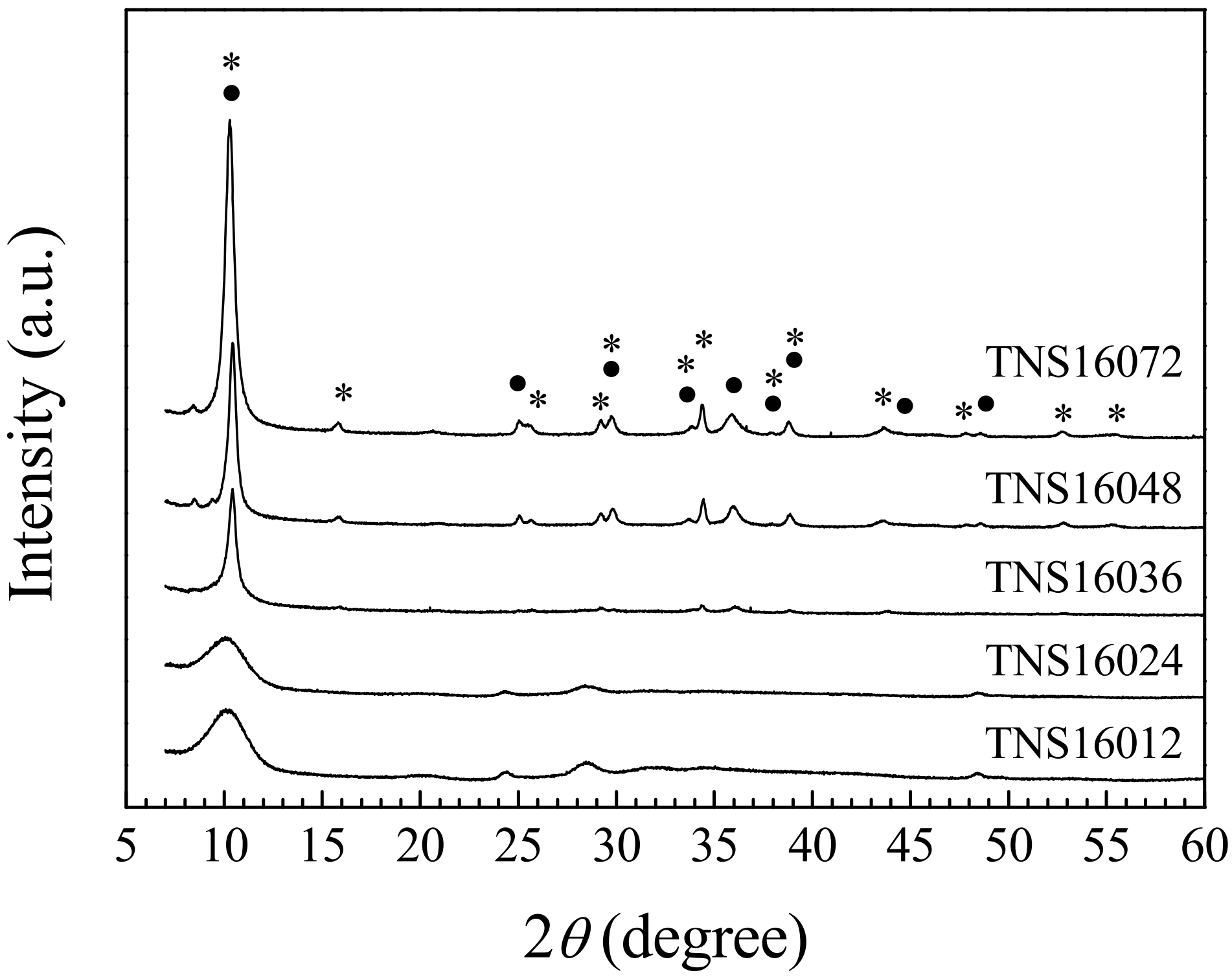

**Figure 3** - XRD patterns of the titanate nanostructures prepared at 160 ºC and different reaction times: 12 h (TNS16012), 24 h (TNS16024), 36 h (TNS16036), 48 h (TNS16048) and 72 h (TNS16072). Symbols: (∗) - $Na_2Ti_3O_7$, JCPDS-ICDD file No. 31-1329, (•) - $H_2Ti_3O_7$, JCPDS-ICDD file No. 41-192.

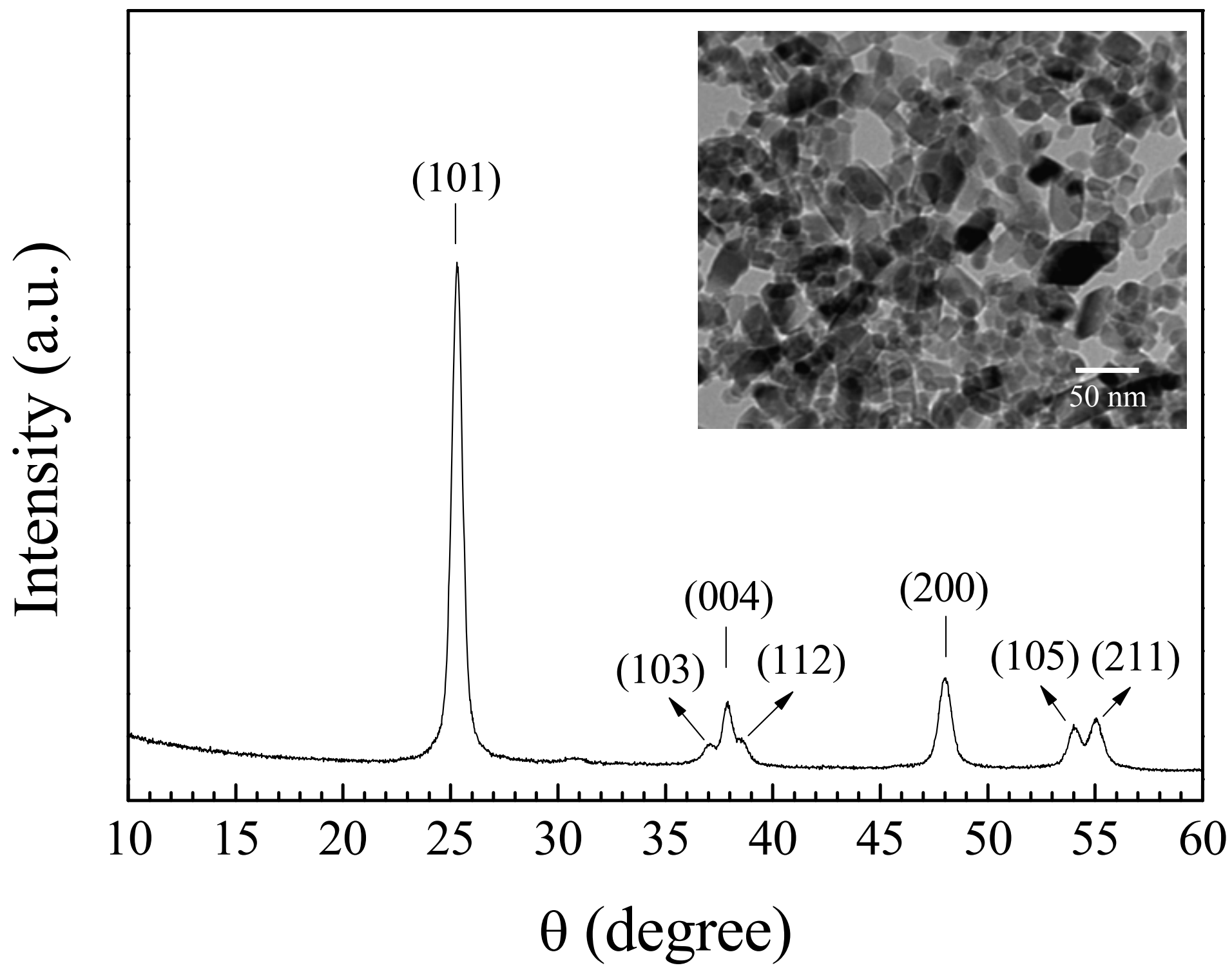

**Figure 4** - XRD pattern of the $TiO_2$ nanoparticles sample NP16024. All the diffraction lines match the $TiO_2$ anatase phase JCPDS-ICDD file No. 21-1272. The inset shows a TEM micrograph of the same sample.

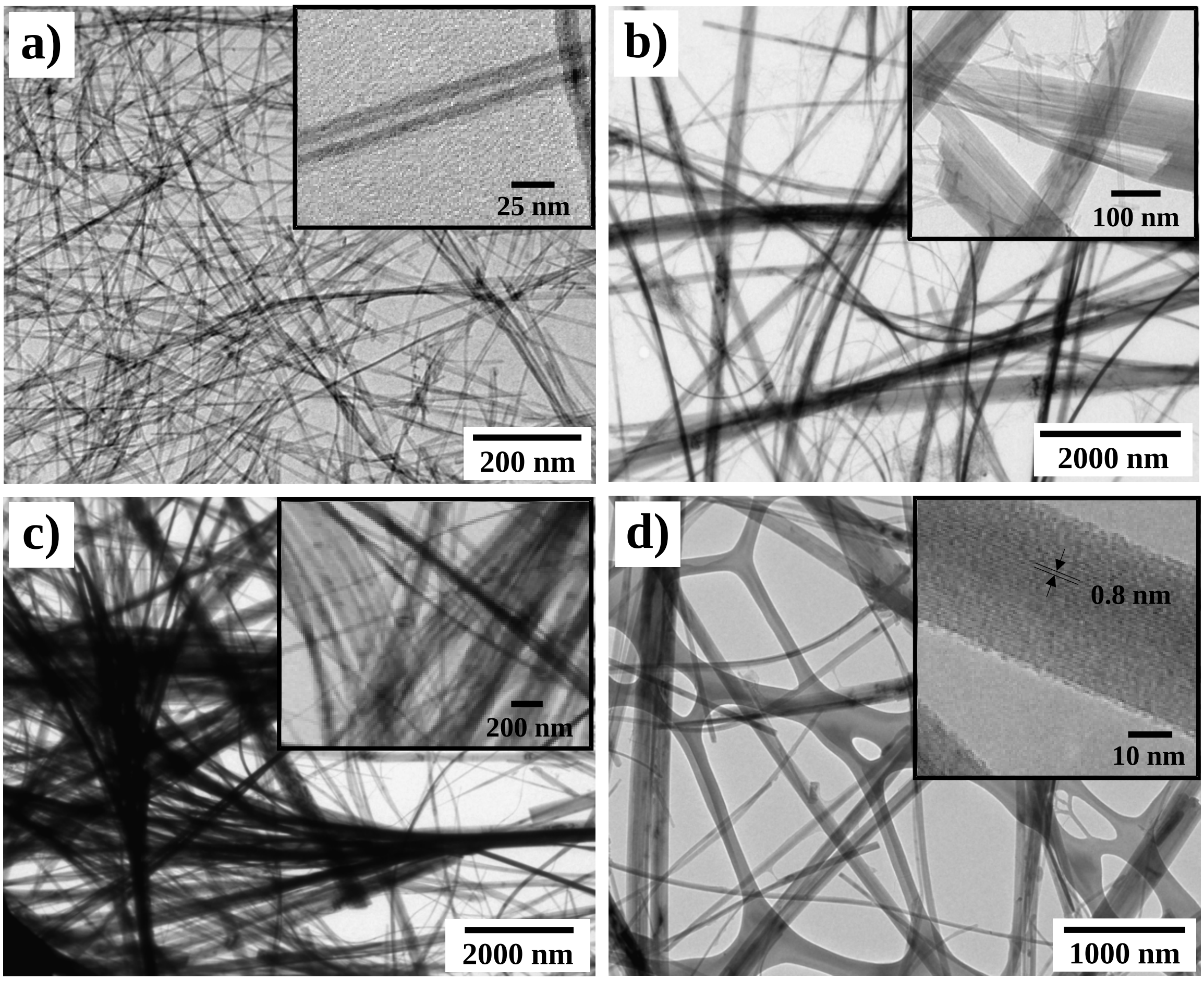

**Figure 5** - TEM images of titanate nanostructures prepared at 160 ºC for a) 24 h, b) 36 h, c) 48 h and d) 72 h.

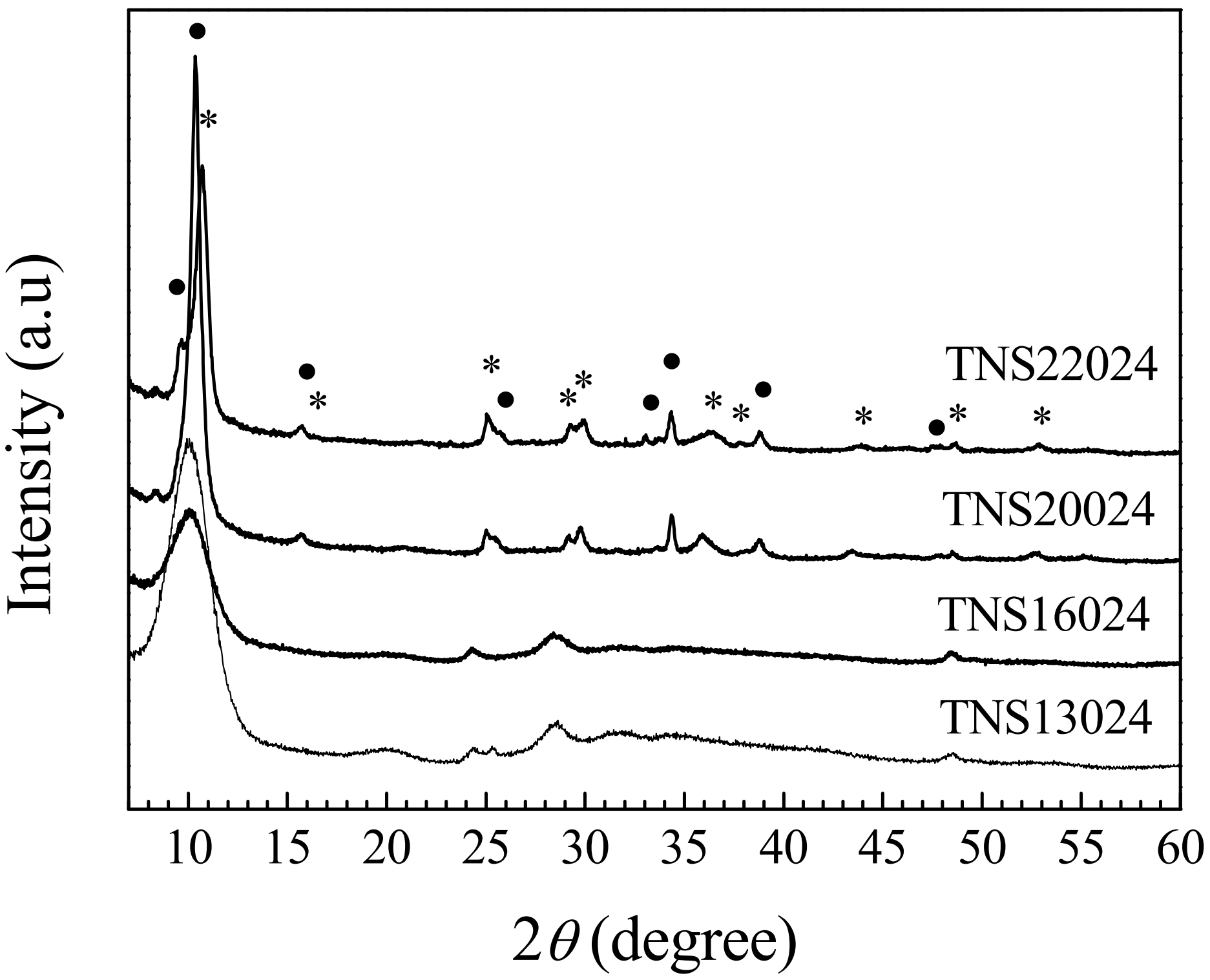

**Figure 6** - XRD patterns of the titanate nanostructures synthesized for 24 h at different temperatures. Symbols: (∗) - $Na_2Ti_3O_7$, JCPDS-ICDD file No. 31-1329, (•) - $H_2Ti_3O_7$, JCPDS-ICDD file No. 41-192.

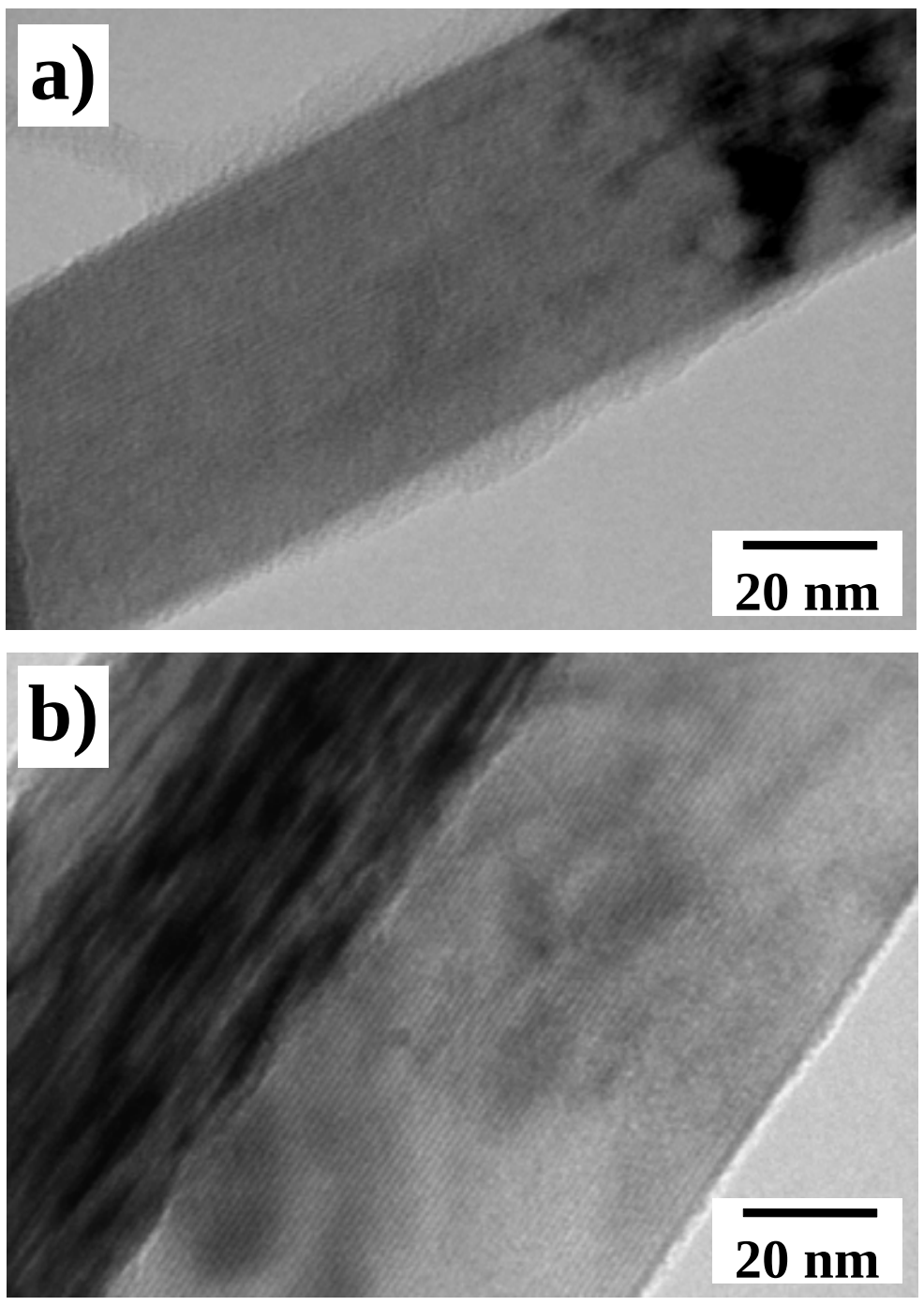

**Figure 7** - TEM micrographs of TNS samples prepared at a) 200 ºC and b) 220 ºC for an autoclave dwell time of 24 h.

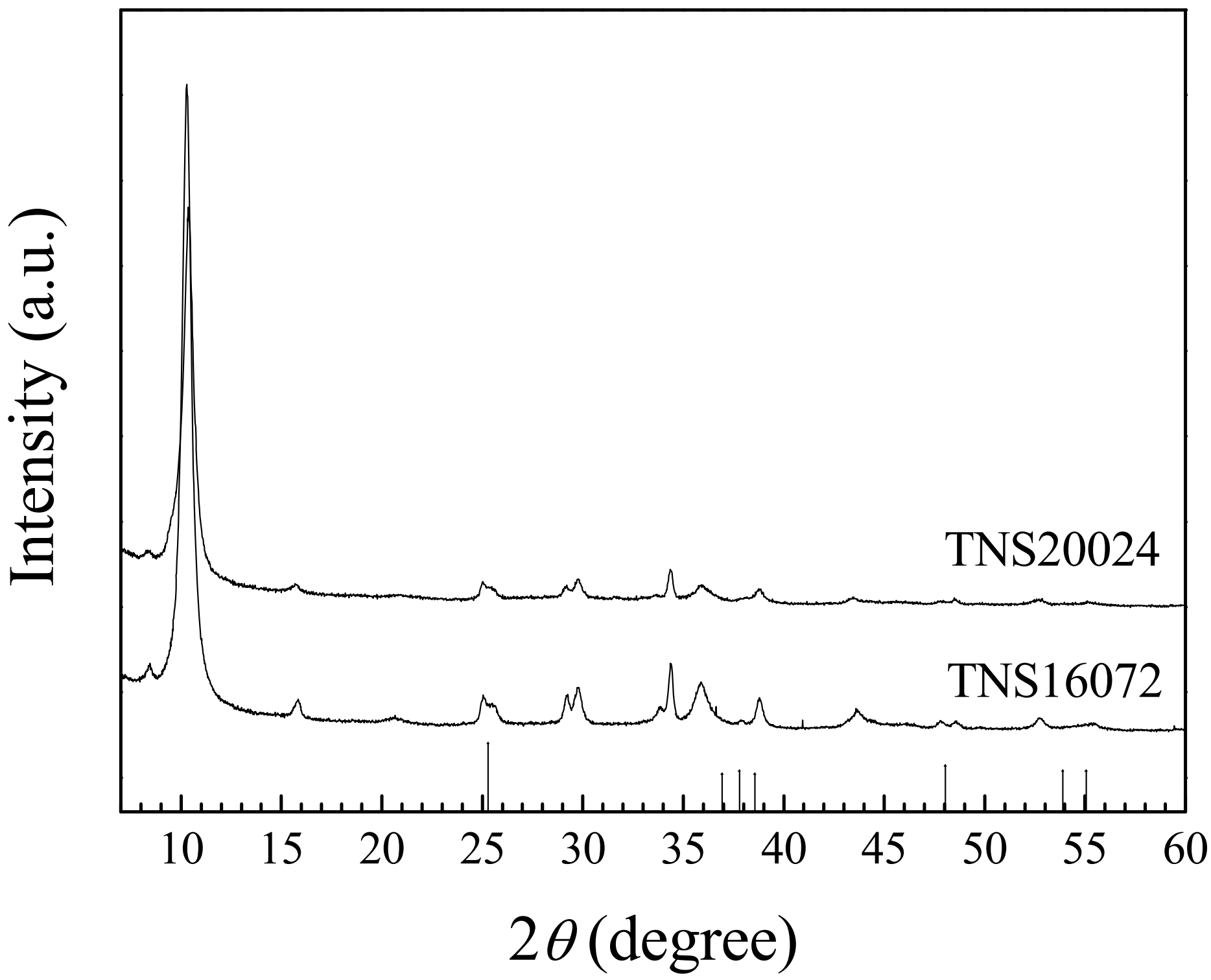

**Figure 8** - XRD patterns of the samples prepared at 200 ºC and 24 h (TNS20024) and at 160 ºC and 72 h (TNS16072). The vertical lines are assigned to the $TiO_2$ *anatase* phase diffraction pattern according to the JCPDS-ICDD file No. 21-1272.

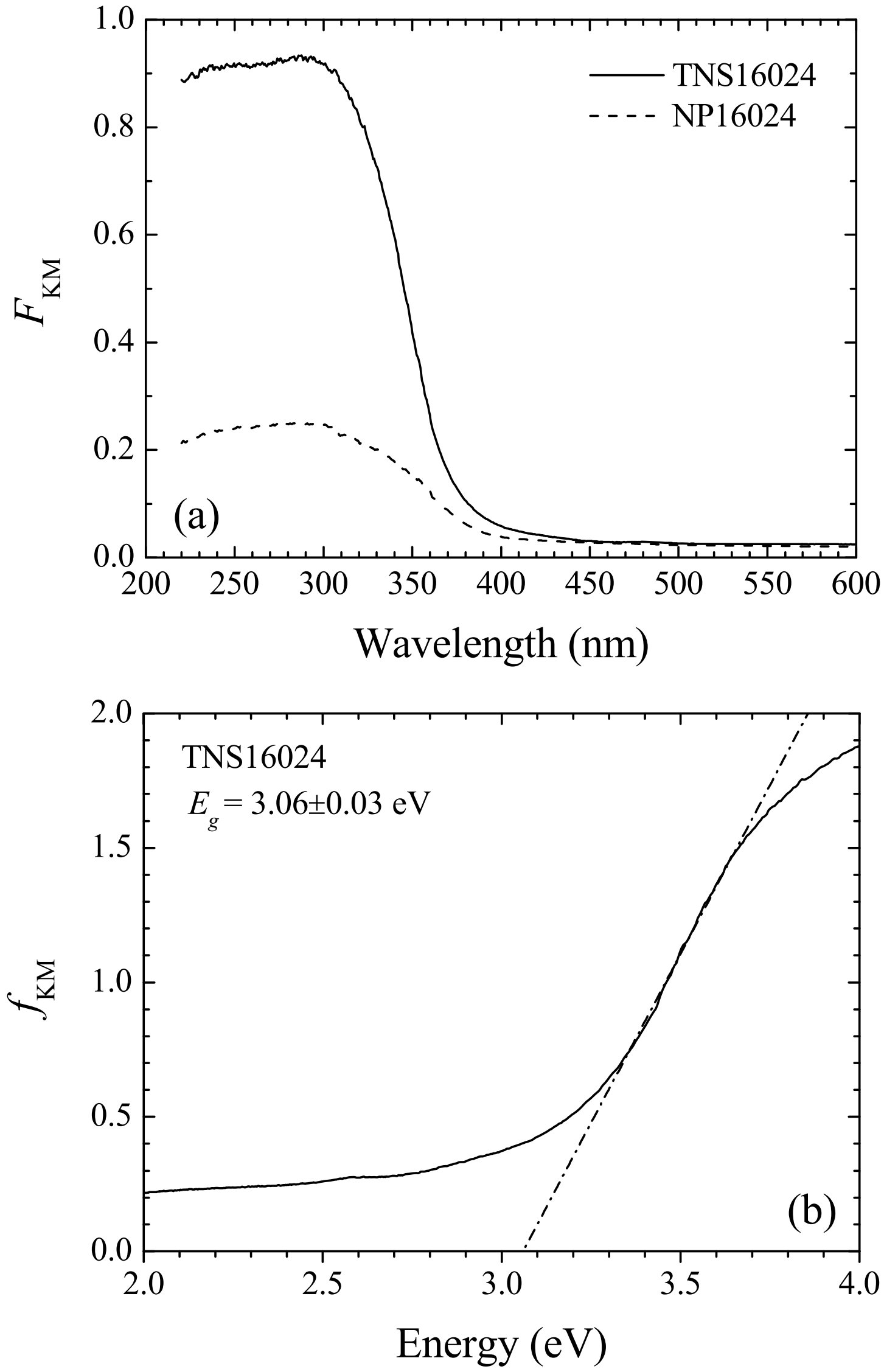

**Figure 9** - a) Diffuse reflectance spectra of the titanate TNS16024 (solid line) and nanoparticles NP16024 (dashed line) samples, prepared at 160 ºC for 24 hours. b) Tauc plot of the TNS16024 sample from which an indirect band gap of 3.06±0.03 eV was determined.

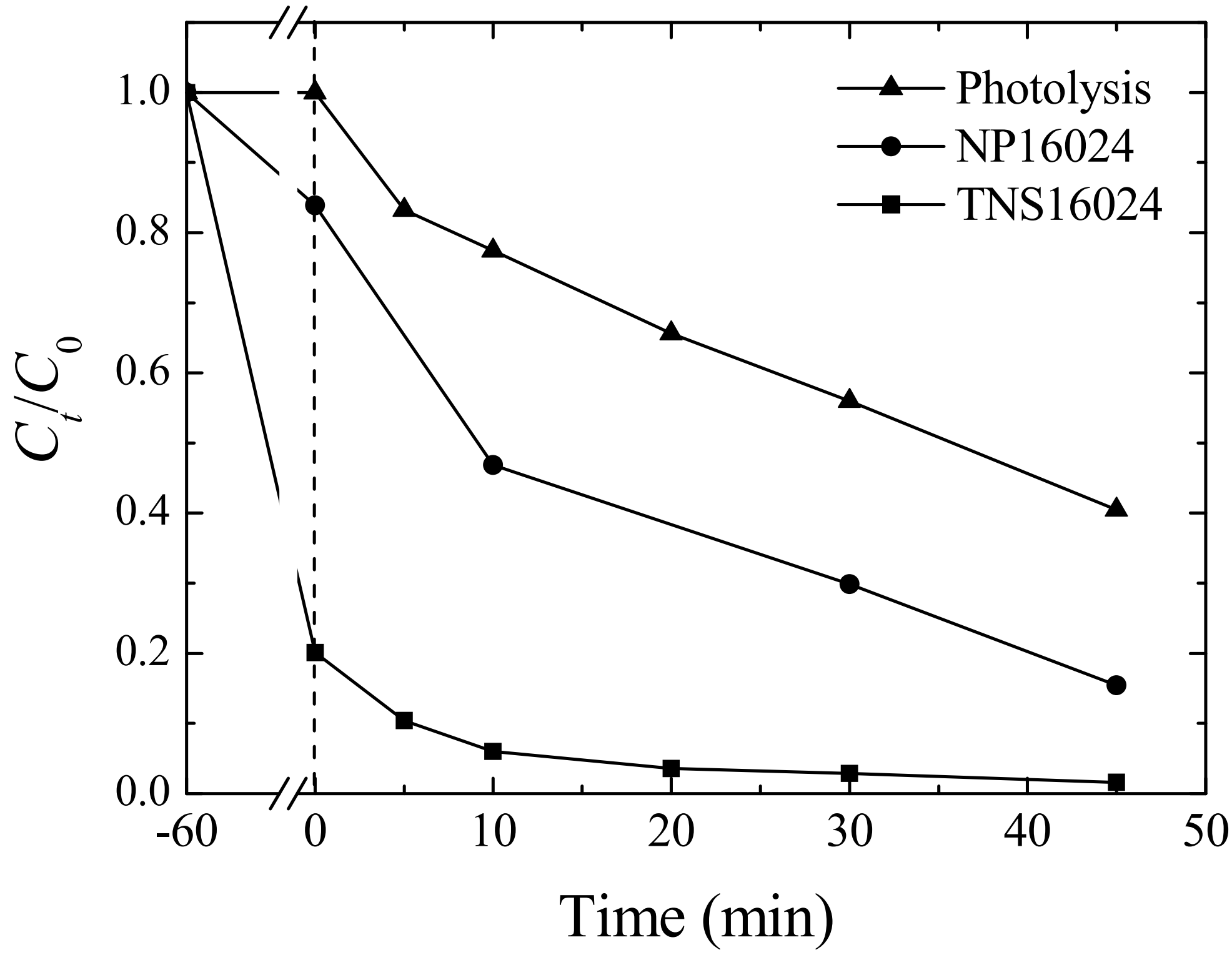

**Figure 10** - Photocatalytic decolourization of a 10 ppm R6G aqueous solution (150 ml) using 15 mg of catalyst. $C_i$ and $C_t$ stands for the initial and over time concentrations of R6G, respectively.